\documentclass[3p]{elsarticle}
\makeatletter
\def\ps@pprintTitle{%
 \let\@oddhead\@empty
 \let\@evenhead\@empty
 \def\@oddfoot{}%
 \let\@evenfoot\@oddfoot}
\makeatother
\usepackage[hidelinks]{hyperref}
\usepackage{amsmath,amsthm,amssymb,bm}

\usepackage{amssymb}
\usepackage[overload]{empheq}
\usepackage{latexsym,dsfont,textcomp}

\usepackage{commath,mathtools,mathrsfs}
\usepackage{parskip}
\usepackage{epstopdf,pdfpages,graphicx}
\usepackage[nottoc]{tocbibind}
\usepackage{array}
\usepackage{babelbib}
\usepackage[english]{babel}

\newtheorem{thm}{Theorem}[section]

\newtheorem{defi}[thm]{Definition}

\graphicspath{ {./images/} }
{
      \theoremstyle{plain}
      \newtheorem{ass}{Assumption}
  }
  
\newcommand{\R}{\mathbb{R}}
\newcommand{\N}{\mathbb{N}}

\newtheorem{innercustomgeneric}{\customgenericname}
\providecommand{\customgenericname}{}
\newcommand{\newcustomtheorem}[2]{%
  \newenvironment{#1}[1]
  {%
   \renewcommand\customgenericname{#2}%
   \renewcommand\theinnercustomgeneric{##1}%
   \innercustomgeneric
  }
  {\endinnercustomgeneric}
}

\newcustomtheorem{axiom}{Axiom}

\begin{document}
\begin{frontmatter}
\vspace{-0.6cm}
\AtBeginEnvironment{quote}{\itshape}

\title{Utilitarianism on the front lines: \\ COVID-19, public ethics, and the "hidden assumption" problem}
\author[1]{Silvio Vanadia}
\author[2]{Charles Shaw}

\address[1]{Independent}
\address[2]{FixedPoint IO}

\begin{abstract}
How should we think of the preferences of citizens? Whereas self-optimal policy is relatively straightforward to produce, socially optimal policy often requires a more detailed examination. In this paper, we identify an issue that has received far too little attention in welfarist modelling of public policy, which we name the "hidden assumptions" problem. Hidden assumptions can be deceptive because they are not expressed explicitly and the social planner (e.g. a policy maker, a regulator, a legislator) may not give them the critical attention they need. We argue that ethical expertise has a direct role to play in public discourse because it is hard to adopt a position on major issues like public health policy or healthcare prioritisation without making contentious assumptions about population ethics. We then postulate that ethicists are best situated to critically evaluate these hidden assumptions, and can therefore play a vital role in public policy debates.

\end{abstract}
\end{frontmatter}

\newpage{  \hypersetup{linkcolor=black}  \tableofcontents}


\newpage

\section{Introduction}
The COVID-19 outbreak, which has resulted in restrictions on transportation, work, and ordinary life, has
affected billions of people and cost the global economy trillions of dollars as a result of a possible hazard
to many lives. It has also stimulated wide-ranging public debate over complex social dilemmas. This is
not surprising because in order for a society to make choices, we need a set of rules or principles to follow.
Medical, governmental, and political choices must be made, understood, and applied. We expect decisions
which affect the public to be grounded in presuppositions which have been consciously chosen and critically
assessed, to a robust philosophical standard. Can we rely on ethics, especially
population ethics, to evaluate the merits and demerits of a government’s response to a pandemic? Whilst
moral and ethical theories are occasionally referenced, the contribution of ethicists to public policy in times
of crisis, including the COVID-19 pandemic, has been unclear.

The aim of this article is twofold. Motivated by the Italian experience, our primary goal is to discuss the
approach that national authorities have taken, as well as the specific value frameworks that they have used
or proposed to establish the priority of treatment of patients in intensive care. We examine the ethical issues
emerging in the context of a public crisis, such as the COVID-19 pandemic.

Our secondary goal is to examine decision-making behind resource allocation and some of the ethical difficulties
that society and healthcare systems face in the context of a public emergency. In regular circumstances,
it is the job of hospital ethics committees to assist in the resolution of ethical quandaries in the setting of
inpatient treatment. Clinicians are accustomed to similar quandaries: setting care objectives, evaluating
patients’ decision-making competence, handling treatment refusals, and so on. Ordinarily, clinicians would
address these quandaries with a patient-centered approach, but these are not ordinary circumstances.

It is not our aim to provide a framework for the resolution of extant moral quandaries or even to provide
a tentative roadmap for doing so. This is not least because the crisis is still ongoing. Instead, we aim to shed light on some of the
practical challenges regarding triage ethics, resource allocation, and medico-ethical decision-making under
crisis.

Discussions on social issues are frequently related to policy options, which necessitate evaluation. What does
it mean to state that one policy option is superior than another one? Principles must be able to balance
gains and losses because policy changes seldom result in benefits for everyone. Some policies result in a
greater population that is less well-off than others. As a result, there is an implicit ethical dimension to
policy decisions.

Welfarist evaluation principles are the primary focus of our analysis. Welfarism is founded on the premise that if two policy scenarios (with the same population) result in everyone getting the same amount of reward, they are both equally beneficial \cite{Sen87}.

In accordance with welfarist ideas, values like individual liberty and autonomy are instrumental and desirable
because they contribute to well-being. As well as this, virtues and fair practices may be useful. Since this is
the case, a holistic approach to well-being is essential. For example, it seems appropriate to focus on lifetime
well-being and include enjoyment; pleasure and the absence of pain; autonomy; liberty; and good health. We presume that people who are fully informed and autonomous have self-regarding preferences
that align with their well-being.

Where should we begin our investigation of the moral underpinnings of lockdown rules in the case of a
pandemic? It is well understood that COVID-19 is an example of a pandemic that has had a significant
impact on global welfare and economic prosperity. On the other hand, there is substantial disagreement on
whether there exists an optimal policy approach when social interaction is crucial to both illness transmission
and economic activity. In the face of excessive demand, health systems have attempted to develop rules in
this vein for which patients should be treated. Enquiring minds may want to know the extent to which broad and intuitive analysis of social policy decisions are made as a result of such rules and the corresponding
outcomes. Whereas self-optimal policy is relatively straightforward to produce, socially optimal policy,
which is arguably of more practical importance, requires a more detailed examination.

The remainder of the paper is organised as follows. Section 2 briefly outlines some of the standard philosophical
and ethical frameworks. Section 3 demonstrates that amidst the public discussion over pandemic response,
one ethical framework – utilitarianism – has been both referenced and criticised. Section 4 introduces a
so-called axiomatic approach, which consists of formulating and solving problems related to social choice,
as well as analysing their underlying axioms. Section 5 describes how, in order to account for the feedback
between the status of the pandemic and people’s behaviour, formal (mostly game-theoretic) models of social
distance have gained ascendancy. Section 6 examines the question of assumptions that appear to be hidden
or implicit in models used to guide epidemiological, economic, or public health related decision making.
Section 7 concludes and provides tentative recommendations for future research.

 \section{Some standard philosophical approaches}
 \label{sec:standard}
 Before we begin our analysis, we will briefly outline some of the standard ethical frameworks as follows.
\begin{itemize}

\item ”Act utilitarianism” demands that each of our individual decisions maximise quantifiable good. By
imposing limits (ostensibly bad), governments have attempted to prevent the spread of the pandemic
(good), enhance public health (good), and save lives (good), maybe at the price of other lives (bad).
It will only be possible to evaluate the strategy and timing of 'applying the brakes' once the pandemic is over. Although empirical calculations may favour isolation in retrospect,
the likely sacrifice of some innocent lives will nevertheless leave some moral concerns unaddressed.

\item ”Rule utilitarianism” advises us to devise rules to act on in specific instances. The use of Quality
Adjusted Life Years (QALYs) in decision making is one such concept. If individuals who die are
extremely elderly, they will not have as great of an impact on the national QALY aggregate as the young and healthy.

\item ”Kantian ethics” dictates that one must never treat other human beings as means to an end but only
as ends in and of themselves. Did, for example, the Italian government follow this rule? This can be
argued in either direction. Yes, since the lives were not sacrificed in one given (possibly isolated) region
for the sake of health advantages elsewhere. No, since the weak were placed in danger to help others.
The Kantian formula leaves considerable room for interpretation.

\item ”Natural law” bans us from violating or disregarding our fundamental human rights, which
are, in the traditional formulation, survival, health, shelter, having and rearing children, and pursuing
knowledge, particularly about God. This method easily supports the government’s determination to
safeguard lives and health. However, since the same family of ideas gives rise to the theory of twofold
impact for exceptions, it is not always apparent how the principles should be implemented.

\item "Moral legalism" is not an ethical system in and of itself, but it influences people’s thinking in several
regions of the globe, including Italy. To be moral, we merely need to follow the rules of the land. This is consistent with the legal positivist jurisprudential credo, which simply holds
that the law is the law because it is the law, and that no further explanation is conceivable or desired.
The argument, on the other hand, is incompatible with natural law theory, which holds that morality
is superior to current law and that certain laws are harmful and should not be observed.

\item ”Virtue ethics”, with roots in its Aristotelian formulation, instructs us to find a balance between doing
too much or feeling too strongly and doing too little or feeling too weakly. This concept works particularly
well in hazardous situations: instead of being cowardly (e.g. too little action) or rash (e.g. too much action) in
response to fear, we should follow the golden mean and be courageous (appropriate action). It does
not, however, give precise guidance for making policy decisions.
\end{itemize}

Now that we have outlined some of the standard approaches relevant to the question at hand, let us continue
with a typical challenge of how to distribute limited resources. Successfully addressing this problem can
be achieved in a variety of ways, such as advocating for the distribution of the items in question to those
who need them, deserve them, or can afford them. In our setting, heterogeneity in choice sets may result in
fundamentally different types of state governance. There is temporal variation of choice sets to consider too,
since the impacts of a choice on resource allocation may be long-lasting or subject to hysteresis. In light of
these considerations, any solution to a given distributive justice issue ought to be welcomed. However, we
will argue that this is a naive view. We will show that in the face of such complex societal issues, ethics
alone (and even ethics in general) may struggle to provide definitive solutions. On the other hand, ethics
has a direct role to play in public discourse because it is hard to adopt a position on major issues like public
health policy or healthcare prioritisation without making contentious assumptions about population welfare.
 
\section{"Ruthless" vs "soft" utilitarianism}
\label{pt:SIAARTI}
Health systems have a responsibility to manage finite resources during a pandemic \cite{E+}. This guideline, widely accepted in countries such as the United States \cite{NY} and the United Kingdom \cite{RCP}, mandates the allocation of resources in order to "maximise the number of patients who survive therapy with a realistic life expectancy." Ethical dilemmas might arise in the process of obtaining such goals. For example, it is "justifiable" to remove a patient from an ICU bed or a ventilator "since maximising benefits is crucial," according to experts. "Patients who are fairly regarded to have the ability to benefit fast should be given access to ventilators," the British Medical Association recommends \cite{BMA}.

Should the optimal strategy, if one exists, be to reduce infections until a vaccine and a treatment are
available, or is herd immunity preferable? When a pandemic shows indications of abating, how quickly
should the limits on economic activity be lifted, either organically or as a result of a vaccination campaign?
Whilst there is considerable uncertainty in both the answer to this question and the factors that affect this
response, moral arguments continue to be shaped by the pandemic. Many of them centre on a more holistic
view of ethics, which emphasises justice and care for the vulnerable. Amidst public discussion regarding the
response to the pandemic, one ethical framework – utilitarianism – has been both referenced and criticised.

In March 2020, the civil rights office of the U.S. Department of Health and Human Services stated that:

\begin{quote}
"persons... should not be put at the end of the line for health services during emergencies. Our civil rights laws protect the equal dignity of every human life from ruthless utilitarianism...HHS is committed to leaving no one behind during an emergency, and this guidance is designed to help health care providers meet that goal."
\end{quote}

Roger Severino, Office of Civil Rights Director, U.S. Department of Health and Human Services, also outlined
that healthcare providers that apply triage protocols may be subject to legal liability as he announced the intention of his office to investigate those who apply them \cite{Fink}.

This view seems to conflict with one proposed by the Italian Society of Anesthesia Analgesia Resuscitation
and Intensive Care (SIAARTI), who in March 2020 (in one of the most serious phases of the Italian emergency)
presented a list of recommendations, offered operational guidelines, and examined ethical concerns
in order to assist clinicians involved in the care of COVID patients \cite{Ri}. It advised making choices in the
best interests of the greatest number of people. This idea has various practical implications, including the
recommendation that resources be directed towards younger patients, who may survive longer following
recovery than older ones. They were also urged to utilise resources on patients for whom they would have
a greater impact and on those for whom the treatment would most likely be shorter (thereby making the
resource available to aid another patient). In practice, it is a utilitarian approach to resource allocation that
aims to maximise the overall number of years of life preserved. It entails making trade-offs between years
of life preserved for some and years forfeited for others. Specifically, SIAARTI stated that:

\begin{quote}
"An age limit for the admission to the ICU may ultimately need to be set. The underlying principle would be to save limited resources which may become extremely scarce for those who have a much greater probability of survival and life expectancy, in order to maximize the benefits for the largest number of people." \cite{Shaw}
\end{quote}

The protocol developed by the Piedmont region's civil protection department goes even further, stating that

\begin{quote}
"The requirements for access to intensive treatment in times of emergency must include age of less than 80 or a score on the Charlson comorbidity Index of less than five." \cite{Shaw}
\end{quote}

 Peterson et al \cite{Peterson} call this a ”soft” utilitarian approach, and point out that it is already being used in areas
of Italian healthcare where resources are severely limited, such as organ transplants. By looking at both
the likelihood of survival and ”maximum life expectancy,” as well as projected length of ICU stay and the
consequent use of intensive care resources, the Italian authorities use this strategy to maximise potential
value in terms of additional years of life \cite{Cillo}. However, some authors have highlighted that this strategy is
in direct contrast with the egalitarianism of the Italian healthcare system\cite{CVSW}.
 
 
Whilst terms such as ”soft” utilitarianism have may sound appealing, they offer limited scope for understanding,
let alone estimating, the welfare loss of a pandemic. During a pandemic, one of the simplest
methods for people to lessen their risk of infection is to limit their contact with infected persons. These
notions do not include social distancing strategies, which are behavioural modifications that reduce contact
rates between those who are at risk of contracting a disease and those who are already ill. The severity
of a pandemic can be reduced by social distancing techniques, but the advantages of social distancing rely
on the extent to which individuals employ it. When people do not want to pay the expenses associated with
social separation as an effective form of control, it can be less successful.

Clearly, such seemingly ad-hoc categorisation of utilitarianism as ”soft” or ”ruthless” lacks applicability.
Should we, for example, interpret ”ruthless” utilitarianism as pure utilitarianism? Probably not. Pure
utilitarianism (the non-rule variety) proves very difficult to apply on the level of policy, as evidenced by the
many paradoxes which arise when it is taken to its logical conclusion. So what kinds of utilitarianism, if they are at
all comparable, did the US authorities find so objectionable and Italian authorities advocate? The Italian
authorities were arguably more explicit with their recommendations.
 
 The SIAARTI recommendations can be summarised as follows \cite{V+a,V+b}:
 
\begin{quote}
\begin{enumerate}
\item When the availability of resources is overwhelmed by their need, a decision to deny access to one or more life-sustaining therapies, solely based on the principle of distributive justice, may ultimately be justified.
\item  Criteria for allocation should be flexible and adapted locally in response to available resources, the potential for patient transfer, and the ongoing or foreseen number of admissions.
\item  An age limit for admission to the ICU may ultimately need to be set.
\item  Together with age, the comorbidities and functional status of any critically ill patient should be carefully evaluated.
\item  Every admission to the ICU should be considered and communicated as an “ICU trial.” The appropriateness of life-sustaining treatments should be re-evaluated daily.
\end{enumerate}    
\end{quote}
The extant literature examining the Italian response seems to focus on the early reactions (disbelief and
inactivity) of both the Italian national and regional administrations, as well as the general population. Later
in the crisis, however, the national government took rapid and decisive action to address the health crisis
because of the early designation of an emergency state. Could Italian authorities have followed China’s
example and implemented much stricter controls? Lockdown measures like those imposed by China in the
middle of the crisis would arguably have been difficult to implement because they necessitated a trade-off
between individual rights and the need to limit, or at least minimise, the spread of the virus, which was
unprecedented in a democratic government. Non-essential production operations would arguably have had
to be shut down as a result of a lack of support from both the business industry and union officials for lockdown
measures \cite{Bosa}.

A wide-ranging debate remains as to whether the Italian government’s reaction, as well as
other variables, played a significant causal a role in the high death rate. Given Italy’s lack of centralization,
readiness and containment may have been compromised. Italy’s age structure, its geographical concentration,
and slow and poor decision-making all contributed to the spread of the virus into Italy’s care-homes and other
vulnerable populations, as did a dearth of knowledge of the virus and its effects, as well as a lack of both national and regional
coordination, which allowed the virus to enter the care-home sector.
Lockdown had become the norm throughout Europe and the rest of the globe by the time Italy started
lifting restrictions in May. As a result, the government was able to maintain control over the virus, and a
considerable period of time (spring and summer) followed before the second wave started to present new
issues.
 
In terms of the "standard framework" of ethics (see section \ref{sec:standard}), the SIAARTI recommendations may plausibly be interpreted as something akin to "rule utilitarianism". Translated into utility theory, it states that although preferences are pre-defined, choices are included in the model\footnote{Which is obviously very different from e.g. categorical imperative where choices are/should be predefined in the model. Needless to say, public policy subscribes to "rule utilitarianism" far more frequently.}. But does it do this in a logically satisfactory way? In other words: does it subscribe to e.g. "rule utilitarianism" in a way that is internally consistent and logically valid? An alternative 
is what Thomson and Lensberg \cite{TL} refer to as the "axiomatic approach".

\section{Axiomatic approach}
 \label{sec:axiom}
The axiomatic approach consists in formulating and then solving problems related to social choice. The problems are then checked to see whether they are compatible: that is, whether or not there are solutions which
fulfil all of the axioms. The axioms are the key intellectual challenge in this framework. The difficulties
arise, inter alia, in the process of achieving internal consistency, a desiderata of formal-axiomatic systems.
For example, the axioms may conflict. The desire to impose more axioms than are mutually compatible is
not unheard of either.

If a solution meets all of the axioms on a one-by-one basis, it may turn out to be unsatisfactory in a specific
context. Then, the logical next step is to identify the class of cases exemplified and create an extra axiom
indicating how the solution should behave on that class, and lastly to discover the biggest subset of axioms
from the original list that are consistent with the new axiom. There may be several unique subsets that
are compatible with each other. Maximal compatibility analysis reveals the ”cost” of each axiom and the
”trade-offs” between axioms through methodical research. Ultimately, progress in addressing the problems
of social choice may be achieved on the basis of such facts and one’s intuition about the various axioms. 

This allows for a richer and more concrete taxonomy, such as one proposed by Blackorby et al \cite{BBD}and includes
critical-level generalized utilitarianism, restricted critical-level generalized utilitarianism, number-sensitive
critical-level generalized utilitarianism, restricted number-sensitive critical-level generalized utilitarianism,
umber-dampened generalized utilitarianism \cite{Hurka}, and rank-discounted critical-level generalized utilitarianism \cite{AZ,Spears}. The main properties of the axiomatic population ethics framework are tabulated in Table \ref{tab:principles}, which is a non-exhaustive list. 

\begin{table}[htb!]
    \centering
\begin{tabular}{lllllll}
   &       & Utility      & Existence    & Negative     & Avoidance of the      & Priority for       \\
   &       & Independence & Independence & Expansion    & Repugnance Conclusion & Lives Worth Living \\ \hline
1  & CU    & $\checkmark$ & $\checkmark$ & $\checkmark$ & $\checkmark$          &                     \\
2  & CLU   & $\checkmark$ & $\checkmark$ & $\checkmark$ & $\checkmark$          &                    \\
3  & RCLU  &              &              & $\checkmark$ & $\checkmark$          & $\checkmark$       \\
4  & NCLU  & $\checkmark$ &              & $\checkmark$ & $\checkmark$          &                    \\
5  & RNCLU &              &              & $\checkmark$ & $\checkmark$          & $\checkmark$       \\
6  & AU    &              &              &              & $\checkmark$          & $\checkmark$       \\
7  & RAU   &              &              & $\checkmark$ & $\checkmark$          & $\checkmark$       \\
8  & NDU   &              &              &              &                       & $\checkmark$       \\
9  & RNDU  &              &              &              &                       & $\checkmark$       \\
10 & NDU   &              &              &              &                       & $\checkmark$       \\
11 & RNDU  &              &              & $\checkmark$ & $\checkmark$          &                    \\
12 & NDU   &              &              &              & $\checkmark$          & $\checkmark$       \\
13 & RNDU  &              &              & $\checkmark$ & $\checkmark$          & $\checkmark$       \\
14 & RDCLU &              &              & $\checkmark$ & $\checkmark$          & $\checkmark$       \\
\end{tabular}
    \caption{Principles of population ethics \cite{BBD}}
    \label{tab:principles}
\end{table}

The abbreviations in Table \ref{tab:principles} can be explained as follows:

\begin{enumerate}
\item[1] Classical utilitarianism (CU)
\item[2] Critical-level utilitarianism: positive critical level.
\item[3] Restricted critical-level utilitarianism: positive critical level parameter
\item[4] Number-sensitive critical-level generalized utilitarianism
\item[5] Restricted number-sensitive critical-level generalized utilitarianism: restricted version of 4
\item[6] Average utilitarianism (AU)
\item[7] Restricted average utilitarianism
\item[8] Number-dampened utilitarianism: general case
\item[9] Restricted number-dampened utilitarianism: general case
\item[10] Number-dampened utilitarianism: ratio of critical level to average utility is a positive constant less than one
\item[11] Restricted number-dampened utilitarianism: restricted version of 10
\item[12] Number-dampened utilitarianism: ratio of critical level to average utility is positive, nondecreasing, and approaches one as population size increases
\item[13] Restricted number-dampened utilitarianism: restricted version of 12
\item[14] Rank-discounted critical-level generalized utilitarianism
\end{enumerate}

For formal definitions of all SWO functions in Table \ref{tab:principles} the interested reader is referred to \cite{TL, BBD, AZ} and references therein. In the below illustrative examples, we provide some formal definitions and state some important conditions. We follow Asheim and Zuber's \cite{AZ} notation.

Let $\textbf{X} = \bigcup_{n \in N} \R^n$ be the set of possible finite allocations of lifetime well-being. For each $n\in \N$, each allocation $n\in \R^n$ determines the finite population size, $n(\textbf{X}) = n$, and the distribution of well-being, $\textbf{x}=(x_1,\ldots, x_{n(\textbf{x})}$
A social welfare relation (SWR) on the set $X$ is a binary relation $\succsim$, where for all $\textbf{x},\textbf{y} \in \textbf{X}, \textbf{x}\succsim \textbf{y}$ implies that the allocation $\textbf{x}$ is deemed socially at least as good as $\textbf{y}$. A complete, reflexive, and transitive SWR is called a social welfare order (SWO). For each $\textbf{x} \in \textbf{X}$, let $\textbf{x}_{[~]}=x_{[1]},\ldots, x_{[r]}, \ldots x_{[n\textbf{x}]}$ denote the nondecreasing allocation that is a reordering of $\textbf{x}$. 

The following axioms are uncontroversial in population ethics.

\begin{axiom}{1}[Order]
\label{A1}
The relation $\succsim$ is complete, transitive, and reflexive on $\textbf{X}$.
\end{axiom}

\begin{axiom}{2}[Continuity]
\label{A2}
For all $n \in \N$ and all $\textbf{x} \in \R^n$, the sets ${\textbf{y} \in \R^n : \textbf{y} \succsim \textbf{x}}$ and ${\textbf{y} \in \R^n : \textbf{x} \succsim \textbf{y}}$
\end{axiom}

\begin{axiom}{3}[Suppes-Sen]
\label{A3}
For all $n \in \N$ and all $ \textbf{x},\textbf{y} \in \R^n $, if $\textbf{x}_{[~]} > \textbf{y}_{[~]}$ then $\textbf{x} \succ \textbf{y}$.
\end{axiom}

\begin{axiom}{4}[Existence independence of the best off]
\label{A4}
For all $n \in \N$ and all $ \textbf{x,y} \in \R^n $, and $z\in \R$ satisfying $z \geq \max \big\{ x_{[n]},\textbf{y}_{[n]} \big\}, (\textbf{x},z)\succsim (\textbf{y},z) $ if and only if $\textbf{x} \succsim \textbf{y}$.
\end{axiom}

\begin{axiom}{5}[Existence independence of the worst off]
\label{A5}
For all $\textbf{x,y} \in \textbf{X} $, and $z\in \R$ satisfying $z \geq \min \big\{ x_{[1]},\textbf{y}_{[1]} \big\}, (\textbf{x},z)\succsim (\textbf{y},z) $ if and only if $\textbf{x} \succsim \textbf{y}$.
\end{axiom}

\begin{axiom}{6}[Existence of a critical level]
\label{A6}
There exists $c \in R_{+}$ and $n \in \N$ such that, for all $x \in R^n$ satisfying $x_{[n]} \leq c, (\textbf{x},c) \sim \textbf{x}$
\end{axiom}

All axioms above are also satisfied by ordinary critical-level generalized utilitarianism. However, as Asheim and Zuber's \cite{AZ} points out, the CLU SWO yields the Repugnant Conclusion when $c=0$ and leads to the Very Sadistic Conclusion if $c>0$. The authors propose a weaker axiom as follows and argue that axiom \ref{A7} is key to avoiding the Repugnant and Very Sadistic Conclusions, while not by itself contradicting these conclusions:

\begin{axiom}{7}[Existence of egalitarian equivalence]
\label{A7}
For all $\textbf{x,y} \in \textbf{X}$, if $\textbf{x} \succsim \textbf{y}$, there exists $z \in \R$ such that, for all $N \in \N, \textbf{x} \succ (z)_n \succ \textbf{x}$ for some $n \geq N$
\end{axiom}

\begin{axiom}{8}[Existence of egalitarian equivalence]
\label{A8}
For all $n,m \in N$, for all $x,y \in \R^n$, and for all $\textbf{u,v} \in \R^m, (\textbf{y,u}) \succsim (\textbf{x,v}) \Longleftrightarrow (\textbf{x,v}) \succsim (\textbf{y,v})$ 
\end{axiom}

It is possible to define established generalized utilitarian SWOs of population ethics. 

\begin{defi}[AU]
\label{AU}
An SWR $\succsim$ is an average utilitarian\footnote{also called average generalized utilitarian} (AU) SWO is there exists a continuous and increasing function $u: \R \implies \R$ such that for all $\textbf{x,y} \in \textbf{X}$,
$$
\textbf{x} \succsim \textbf{y} \Longleftrightarrow 
\frac{1}{n(\textbf{X})} \sum^{n(\textbf{X})}_{r=1} u(x{[r]}) \geq 
\frac{1}{n(\textbf{Y})} \sum^{n(\textbf{Y})}_{r=1} u(Y{[r]})
$$
\end{defi}

\begin{defi}[CLU]
\label{CLU}
An SWR $\succsim$ on $\textbf{X}$ a critical-level generalized utilitarian (CLU) SWO if there exist $c \in \R_{+}$ and a continuous and increasing function
$$ 
\textbf{x} \succsim \textbf{y} \Longleftrightarrow  
\frac{1}{n(\textbf{X})} \sum^{n(\textbf{X})}_{r} u(x{[c]}) \geq 
\frac{1}{n(\textbf{Y})} \sum^{n(\textbf{Y})}_{r} u(Y{[c]})
$$
\end{defi}

The CLU SWO with $c = 0$ is the total generalized utilitarian (TU) SWO \cite{AZ}. CLU SWO with $c >0$ also leads to the Very Sadistic Conclusion that there is an egalitarian allocation with positive well-being that is worse for every allocation with negative well-being. The RDCLU SWO may be attractive since it offers an escape from the Repugnant and Very Sadistic Conclusions.

\begin{defi}[RDCLU]
\label{RDCLU}
An SWR $\succsim$ on $\textbf{X}$ is a rank-discounted critical-level generalized utilitarian
(RDCLU) SWO if there exist $c \in \R_{+}, \beta \in (0,1)$ and a continuous and increasing function $u: \R \implies \R$ such that for all $\textbf{x,y} \in \textbf{X}$,
$$ 
\textbf{x} \succsim \textbf{y} \Longleftrightarrow 
\frac{1}{n(\textbf{X})} \sum^{n(\textbf{X})}_{r=1} \beta^r (u(x{[r]})-u(c))  \geq 
\frac{1}{n(\textbf{Y})} \sum^{n(\textbf{Y})}_{r=1} \beta^r (u(Y{[r]})-u(c))
$$
\end{defi}
According to the generalised utilitarian criteria, the function u transforms lifelong well-being into transformed
values. The term ”rank-discounted” refers to the fact that the utility weights are discounted according
to rank by a geometrically decaying function, rather than just being rank-dependent. This is similar to
the time-discounted utilitarian criteria of intertemporal social choice, which uses ”time-discounted utilities”
rather than ”time-dependent utilities.” The parameter c is referred to as a critical-level parameter. The
recognised generalised utilitarian SWOs of population ethics can now be defined. To be consistent with the
literature on intertemporal social choice, we refer to the converted values as utility and $\beta$ as a rank utility discount factor. 

Axioms \ref{A1}-\ref{A7} produce Rank-discounted critical-level generalized utilitarianism (RDCLU)\footnote{This is achieved via Lemmas 1–4 in \cite{AZ},  proof is presented in \cite{AZ}, page 642}. Similarly, axioms \ref{A1}-\ref{A6} together with axiom \ref{A8} produce Critical-level generalized utilitarianism (CLGU).  
Spears and Zuber \cite{SZ} present new axiomatic characterizations of utilitarian (that is, additively separable) social welfare functions that generalize Blackorby et al.'s \cite{BBD98} Expected Critical-Level Generalized Utilitarianism (ECLGU). Spears et al \cite{Spears} further generalise the results of Asheim and Zuber's \cite{AZ} and explain that those who are drawn to Axioms 1-6 but wish to keep the common policy-evaluation practice of same-number independence might be tempted to use CLGU, which incorporates:
\begin{enumerate}
    \item priority for the worse-off (for concave $u$), and 
    \item utilitarianism as special case (for linear $u$).
\end{enumerate}

More recently, Pestieau and Ponthiere \cite{Pest} provide a discussion of the Repugnant Conclusion in the context of the current COVID-19 pandemic and the design of an optimal lockdown policy. They propose that some forms of utilitarianism lead to maximum containment (albeit under certain conditions): for each lockdown with low-quality life periods, there must be a harsher lockdown that is considered as superior, even if it lowers the quality of life periods even more.

\section{Game-theoretic models}
 \label{sec:game}
In order to account for the feedback between the status of the pandemic and people’s behaviour, formal (i.e.
game-theoretic) models of social distance have gained ascendancy. Despite their differences, these models
tend to have a similar message: the outbreak is not as serious as expected without taking behavioural
response into consideration. Individuals tend to return to their old habits when the
pandemic subsides and there is less chance of exposure. Due to self-limiting feedback, the pandemic can
persist for a long time in an intermediate-severity state. This is not so awful that all individuals take it
seriously enough to separate themselves, but maintains enough of a threat that some people do.
 
 Game-theoretic study of agents' motivations to adopt precautionary steps to avoid infection during an epidemic was offered in the 1980s by Fine and Clarkson \cite{FC86}. Following Philipson and Posner \cite{PP93} and Kremer \cite{K96}, as well as Geoffard and Philipson \cite{GP96}, the 1990s saw more advanced dynamic analysis. Because of the negative feedback between illness prevalence and the motivation to take precautions, there is a limit to what may be done by voluntary preventative measures. It is impossible to eliminate diseases that spread by chance encounters: for example, the effect of vaccination diminishes as the disease approaches eradication; see, for instance, Geoffard and Philipson \cite{GP97}.
 
 The types of models referenced are well established and have their origins in the early epidemiology literature, such as the seminal studies of Kermack and McKendrick made as far back as 1926. Indeed, many public health professionals use standard epidemiology models of the kind originally introduced by Kermack-McKendrick in 1926 \cite{KM}\footnote{see \cite{A91} for a historical discussion} and recently updated by, inter alia, \cite{AAL, A20, W+20} to study the effects of an optimal lockdown policy in the context of COVID-19. It is possible, using arguments from economics and epidemiology, to investigate the ideal lockdown approach for a social planner who wishes to limit pandemic mortality while reducing lockdown costs to output. To formalise the planner's dynamic control problem, it is common to employ a standard SIR-type (susceptible-infected-removed) epidemiological model and a linear economy, calibrating the model to actual data. The best policy is then determined by the proportion of infected and vulnerable people in the population. The models have certain variations and special cases, such as SI, SIR/SIS/SIRS, SCIR, and SCIS, see also the recent survey article of McAdams (\cite{McA}\footnote{republished as \cite{McA2}, see page 11}. McAdams helpfully presents a tabulated summary of the literature (page 12) -- all from either the economics, epidemiology, or public health literatures. However, none of the papers presented therein satisfactorily address their modelling assumptions. Either the authors make an informal comment about using a "utilitarian welfare function" or, for the most part, they do not make explicit their assumptions. In general, the discussion of welfarist axiology is rare outside of narrowly specialised academic research, despite welfarist assumptions commonly grandfathered in or baked into the policy modelling. 

\section{Probing latent assumptions of policy models}
 \label{pt:models} 
 Let us now turn to the question of assumptions that are hidden or implicit in models used to guide epidemiological, economic, or public health-related decision making. The illustrative example we provide later in this paper, which will undoubtedly be valuable to ethicists and applied philosophers alike, borrows from Alvarez, Argente, and Lippi (2021) \cite{AAL}, a well-cited and recent paper from the economics literature published in one of the world's best known economics journals. In the context of COVID-19, it serves as the foundational study for a line of research articles examining dynamic interventions. Alvarez, Argente, and Lippi rely heavily on Hall et al \cite{Hall}, who use a utilitarian criterion to value the extra years of life lost among those likely to die due to the infection. 

 Next, we will explain how such SIR-type models, composed of three connected nonlinear ordinary differential equations, inform decision-making. Since citizens under lockdown are expected to be inactive, it is common to assume that the goal of a policy planner is to reduce both the present discounted value of deaths and the output costs of lockdown. A common framework for analysis of such questions in epidemiology involves using the SIR epidemiology model, which describes how the virus spreads from infected agents to those who are susceptible, as well as the rates at which infected agents either recover or die. In this framework, a pandemic planner has just one tool at their disposal: the lockdown of citizens. 
 
 It is the planner's dilemma to balance the costs of locking down, which increase as the number of diseased and vulnerable individuals grows, against the costs of a pandemic, which increase as the number of infected and susceptible individuals rises. This framework lends to examination of how the severity and length of a lockdown are affected by the cost of deaths, which is assessed by how much a life is worth statistically, as well as how successful the lockdown is (i.e. how many contacts are reduced when residents are advised to remain at home). Increasing the fatality rate (chance of dying if infected) increases the policy maker’s motivation for lockdown, as is expected to occur after the hospital capacity is surpassed. However, in the absence of non-linear costs, an ideal gradual lockdown may occur, as the costs and benefits of the policy depend on the current state of the system. 

Using quantitative analysis, it is possible to demonstrate these ideas using simulations that enable us to explore
the issue from a practical point of view. Whilst there are many factors to consider while designing the best
lockdown strategy, quantitative findings that draw from such modelling may help to determine which ones are
most significant. Let us illustrate our argument with a fundamental disease transmission model. The model
is composed of three connected nonlinear ordinary differential equations, none of which have an explicit
formula solution. However, basic calculus techniques enable us to extract a wealth of information about
solutions. Along the way, we demonstrate how this basic model contributes to the theoretical underpinning
of public health interventions and illuminates numerous public health foundations. The population at time $t$ may be classified as susceptible to infection $S(t)$, infected $I(t)$, and recovered $R(t)$
\begin{equation}
N(t) = S(t) + I(t) + R(t) \textnormal{ for all } t \geq 0    
\end{equation}
In the above specification:
\begin{itemize}
\item $S(t)$: susceptible persons who are not now infected, but who may become infected in the future. There is a chance that a person who is vulnerable to infection may get infected. Over time, as the virus spreads from its point of origin or as new sources emerge, the number of people at risk will rise (surge period).
\item $I(t)$: persons who have been exposed to the virus. These are people who have previously been exposed to the virus and can pass it on to others. When an infected person is taken out of the contaminated group, he or she might either recover or die.
\item $R(t)$ This is the group of people who have recovered from the virus and are presumed to be immune, $R(t)$ or have died, $D(t)$.
\end{itemize}

Such an analytical framework is beneficial both numerically and qualitatively because the SIR model simplifies many of the difficulties involved with understanding viral dissemination in real time. The population $N$ is normalised to one, hence all findings should be regarded as fractions of the relevant population. The model relies on related ODEs to describe populations. 
The social planner can make a decision to lockdown at fraction $L(t) \in [ 0,\Bar{L}]$ of $S(t) +I(t)$ with effectiveness $\theta \in (0,1)$. Those who are susceptible become infected when they come into contact with $I$ at rate $\beta(1-\theta L(t)^2)$ (a single dot above a letter indicates the first differential with respect to time):
\begin{equation}
\Dot{S}(t)= -\beta S(t) \times I(t) (1-\theta L(t)^2) 
\end{equation}

Among those not under lockdown, $\beta$ is the number of vulnerable agents per unit of time to whom an infected agent can transmit the virus. In the context of this model, it is assumed that the SIR time scale is short enough to ignore births and deaths (except from those caused by the virus) while at the same time allowing for a minimal number of fatalities from the disease. Amongst the infected, a fraction $\gamma$ recovers:
\begin{equation}
\Dot{I}(t) = \beta S(t) \times I(t) (1-\theta L(t))^2 - \gamma I(t)
\end{equation}

A percentage of those infected die at rate of $0 < \phi(I) \leq \gamma$ per unit of time
\begin{equation}
\Dot{N}(t) = D(t) = \phi(I(t)) I(t)
\end{equation}

It is worth noting that, in this epidemiological model, isolating a portion of the population, while economically costly, can be quite effective in reducing the rate at which the vulnerable become infected. This is because the product of the infected and susceptible determines the number of new infections per unit of time. As a result, reducing each person's number of contacts reduces the number of new infections by its square.

Then, a social planner chooses path $\{L(t)\}$ to minimize the present discounted value:

\begin{equation}
V(S,I) = \min_{L(t)} \int_0^\infty e^{-(r+v)t} \Big( \underbracket{w L_t [\tau(S_t+I_t)+1-\tau]}_{\text{GDP loss (t)}}  + \underbracket{\phi(I(t)) I(t) \times [vsl] }_{\text{deaths(t) × value of life}} \Big) 
\end{equation}
by choosing the fraction of population $L(t)$ to lockdown.

According on the parameter $\tau$, testing is either available or not. The parameter $\tau$ therefore corresponds to testing of the recovered i.e. there are two scenarios:

\begin{itemize}
\item with testing of the recovered ($\tau =1$): lockdown to (S+I)
\item without testing of the recovered ($\tau =0$): lockdown to (S+I +R)
\end{itemize}

The social planner solves the Bellman equation, where the value function $V(S,I)$ such that $0 \leq S+I \leq 1$ solves
\begin{align}
(r+\nu) V(S,I) =  &\min_{L \in [0,\Bar{L}]} \Big[\tau (S+I) +1 -\tau \Big] +I \phi(I) \Big[\frac{w}{r} +\chi\Big] + \\
                  & + [ \beta S(t) (1-\theta L(t)) I(t)(1-\theta L(t))] \partial_S V(S,I) ] + \\
                  & + [ \beta S(t) (1-\theta L(t)) I(t)(1-\theta L(t)) - \gamma I(t)] \partial_i v(S,I)
\end{align}
Here, $V (S, I)$ might be read as the minimal predicted discounted loss in output units for implementing the optimal strategy. After setting, the initial conditions $N(0)=1$, $I(0)= \epsilon$, $S(0)=1-\epsilon$, and boundary conditions:
\begin{itemize}
\item Boundary condition at $I = 0$, $\forall S \in (0, 1): V(S, 0) = 0$
\item Boundary condition at $S = 0$, $\forall I \in (0, 1): V(0,I) = 0 = vsl \times \Big(\frac{\phi \gamma}{r+\kappa+\gamma I} +\frac{\phi \gamma}{r+\nu+2\gamma}  \Big) \times I $
\end{itemize}

Data from the World Health Organization, alongside key economic variables, is used to parameterise the model. Whilst our methodology is consistent with \cite{AAL}, we calibrate the parameters to Italian data where practicable.  

Even if a person has already recovered from the illness, the lockdown is still in effect. In this scenario, locking down the population is less efficient since the recovered are also locked down, with the cost of limiting productivity without the advantage of reducing viral transmission. The lockdown effectiveness decreases significantly in the event of no testing compared to the benchmark condition. In both circumstances, the lockdown costs the same amount of time in terms of lost productivity, because the lockdown length is shorter in the absence of testing, but the lockdown affects a higher percentage of workers (recovered agents are also in lockdown). It is the most important element of a situation when a test is unavailable that the lockdown comes to an abrupt conclusion sooner. As time goes on, the percentage of people who have recovered grows, and therefore the lockdown becomes less effective at stopping the virus from spreading by locking down an increasing number of those who do not have it.

A natural question that arises is how to set the fatality rate. Extensive testing has been done in two cases: the Diamond Princess cruise ship (which yields an age-adjusted fatality rate) and the Italian city of Vo' Euganeo (which yields a lower bound mortality rate). Consistent with both cases, we set the mortality rate $\phi = 0.01 \times \gamma$. We then set $\kappa = 0.05 \gamma$ so that the fatality rate is 3\% when 40 \% of the population is infected. The discount factor used by the social planner is set at 5\% annual interest. The likelihood $\nu$ per unit of time that a vaccine or treatment will be discovered means that these medical breakthroughs take an average of one and a half years to become available. In accordance with \cite{AAL}, we normalise output $w=1$. In accordance with \cite{HJK}, we chose the benchmark value for the additional cost of dying $\chi$ to 0. A utilitarian criterion is used by these authors to value the additional years of life lost by individuals who are more likely to die as a result of the illness, and they arrive at a cost of roughly 30 times annual consumption per capita, which is fairly close to our benchmark value of 20 times annual GDP. We set $\chi = 0$ as the penalty death threshold, which is consistent with \cite{HJK} (see \cite{AAL} for a brief discussion).

\begin{figure}[htbp!]
    \centering
\includegraphics[width=0.55\linewidth]{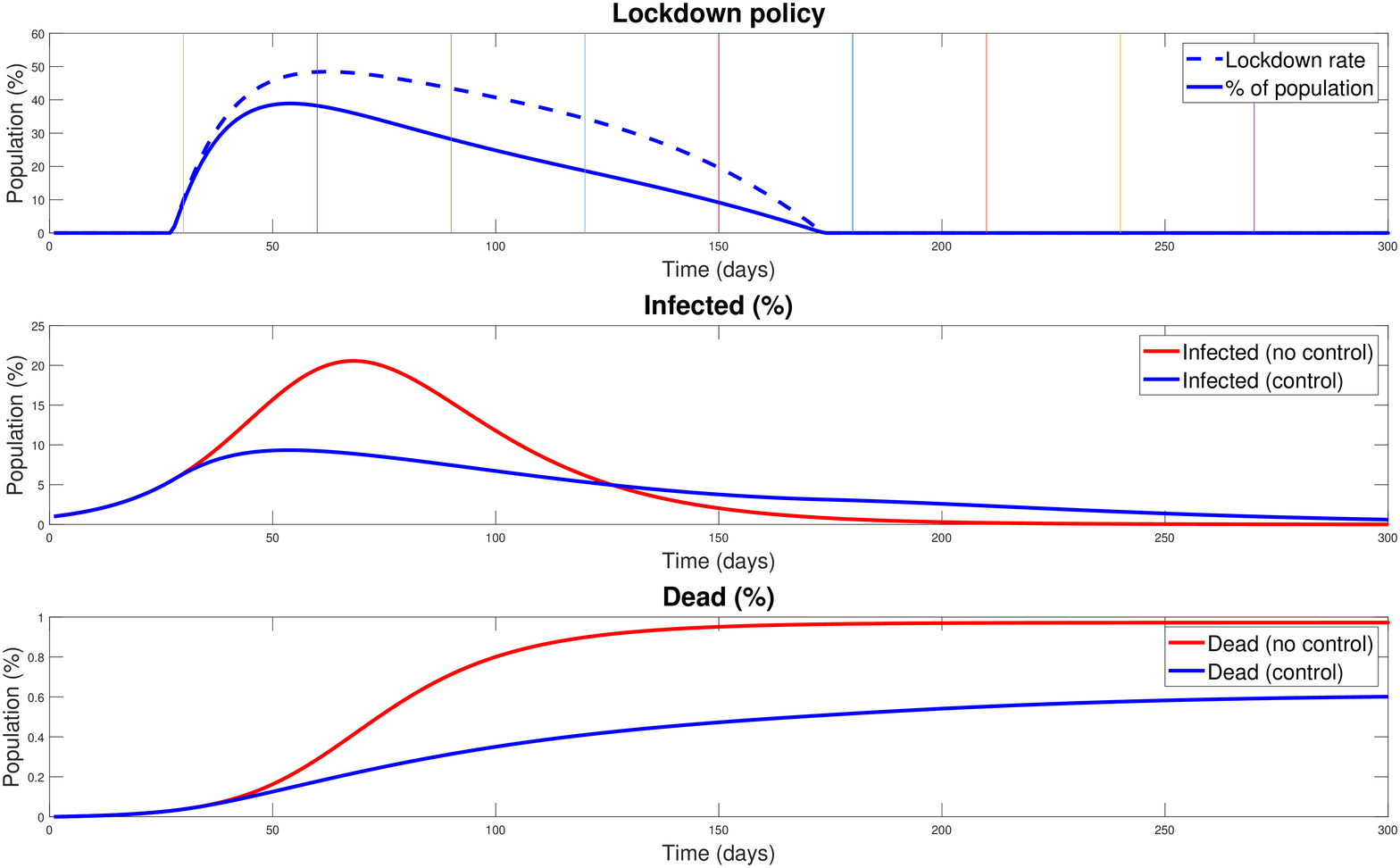}
    \caption{Benchmark Case (medium effectiveness $\theta$ = 0.5)}
    \label{fig:1}
\end{figure}

Figure \ref{fig:1} graphically shows that total welfare costs are almost three times greater due to the cost of deaths. The top panel shows the time paths for the lockdown rate (dashed blue line) and the share of the population under lockdown (solid blue line). The panel in the center shows the share of the population that is infected with control (blue) and without control (red). The bottom panel shows the share of the population that dies under control (blue) and under no control (red). The paths shown are under medium effectiveness (i.e. $\theta$=0.5). The best strategy reduces the overall number of deaths by around 0.80\% in the long run.

\begin{figure}[htbp!]
    \centering
\includegraphics[width=0.55\linewidth]{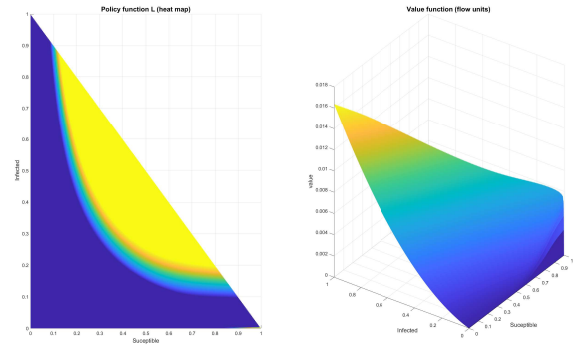}
    \caption{Value Function and Optimal Policy, benchmark case}
    \label{fig:2}
\end{figure}

The value function and optimal policy for the benchmark parameter values are shown in Figure \ref{fig:2}. The value function is represented in the right panel for the appropriate state space (S, I) and normalised as described above, so the vertical axis displays $rV (S, I)/w$. The units represent the cost of permanent flow as a percentage of total output prior to the virus. As a result, a value of 0.02 represents a cost comparable to a permanent reduction of 2\% in the value of production compared to the flow of output before the virus. The best strategy for the benchmark parameter values is shown on the left. Lower lockdown values are indicated by blue, while greater lockdown values are indicated by yellow. The value function is depicted on the left. The units of the value function are permanent flow cost as a percentage of total production prior to the pandemic.

\begin{figure}[htbp!]
    \centering
\includegraphics[width=0.55\linewidth]{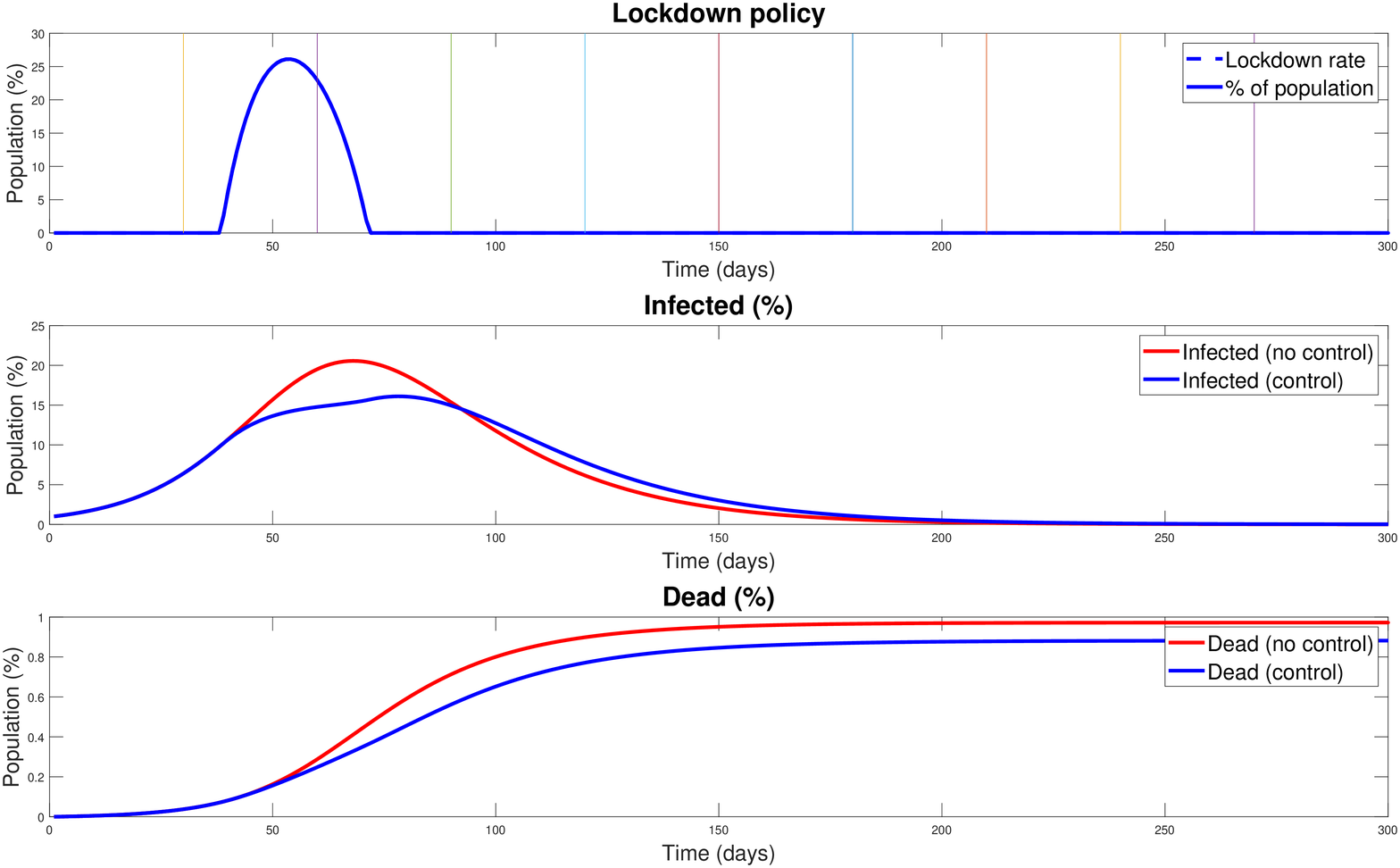}
    \caption{No testing of those recovered ($\tau$ = 0)}
    \label{fig:3}
\end{figure}

Figure \ref{fig:3} demonstrates the welfare effect of testing. The top panel shows the time paths for the lockdown rate and the share of the population under lockdown, which coincide in this case, since there is no testing. The panel in the center shows the share of the population that is infected with control (blue) and without control (red). The bottom panel shows the share of the population that die under control (blue) and under no control (red).\footnote{The case shown is with no testing for those recovered. Numerical results, whilst calculated, are not presented due to being outside of scope of discussion. Full results are available on request.}
 
The authors \cite{AAL} outline some of the assumptions made in the modelling. These are done in order to make the model tractable.

\begin{ass}
The first assumption is that a lockdown is limited to $\Bar{L}\leq 1$. This is due to the fact that some sectors, such as health and fundamental services, cannot be shut down even in the most severe lockdown. 
\end{ass}

\begin{ass}
Members of the population who are sick but not in lockdown are assumed to be able to work as much as those who are susceptible or have recovered. After locking down those exhibiting symptoms, the rest of the group should be isolated. 
\end{ass}

\begin{ass}
Population that is put under lockdown will not produce. 
\end{ass}

\begin{ass}
Except for risk of dying following virus contraction, we assume that all agents are infinitely long-lived. Given the limited time frame of the problem, this simplification is acceptable. However, the lack of a defined age structure results in unrealistic impacts on mortality risk.
\end{ass}

\begin{ass}
The initial proportion of the population susceptible is assumed to be 98\%, or S(0) = 0.98
\end{ass}

Whilst these assumptions are made explicit by the authors, there is a more nuanced (and hidden) assumption which is grandfathered from Hall, Jones, and Klemow (2020) \cite{HJK}. Herein lies our "hidden assumptions" problem. In this example, Hall, Jones, and Klemow use a (classic) utilitarian criterion to value the additional years of life lost by individuals who are more likely to die as a result of the virus.

 Whether philosophers use them, understand them, or appreciate their role in informing public policy-making and advising scientific bodies, such models are important tools of public policy (see, for example, \cite{SIR} and the references cited therein). SIR-type models are the workhorse tools of the COVID-19 pandemic. These models are being utilised by public health experts and policy makers to develop scenarios that are used to guide judgments about recommending or mandating even more stringent mitigation measures on economies (globally) in response to its fast growth. Ethicists and moral philosophers operate at a distance from the front lines. 

However, ethicists and moral philosophers could have an important role to play in the policy debate by offering principled,
philosophical guidance on how to address the extent and bounds of individual liberty in the context of
a pandemic; how to think about collective interests constraining individual interests; and how to assess the soundness of assumptions that underlie mathematical models used in policy modelling. But because
philosophers appear to be largely unfamiliar with these models – along with the assumptions underlying them -
they are not fully involved in the policy debate over social distancing measures. To understand the trade-off
between public health and the policy implications of these mitigation and social distance measures, there is a
case to be made that philosophers ought be aware of the logical consequences of these models. It should be
stressed, however, that such models are particularly sensitive to seemingly innocuous assumptions regarding
the dynamics of infectious diseases.

What other insights, if any, are we able to glean from economics, or one of its sub-fields, such as economic
epidemiology? Arguably, economic analysis has much to offer to the extent that we envisage avenues for
further research where applied philosophy and economics are seen as complements rather than substitutes.
For example, the literature on welfare economics is useful for elucidating the externality elements of infectious
diseases as well as broad policy options. The assumptions about the social health planner, on the other hand,
are accompanied by a set of well-known objections. These include concerns regarding knowledge of the social
welfare function, the availability of proper incentives which results in resource allocation that maximises
social welfare, and the lack of unexpected consequences that weaken welfare efforts. The traditional method
places the social health planner outside of society and separates policy-making from economic analysis.

Nevertheless, there are some meaningful insights to be obtained. Whilst the literature is both vast and
actively developing, we point the interested reader to the recent work by Toxvaerd and Rowthorn [55] who,
building on the economic epidemiology literature, compare socially optimum treatment and immunisation
strategies to the old non-economic approach. Non-economic policies not only have poorer social welfare than
socially optimal policies, but they may even be outperformed by fully non-responsive policies that do not
maximise any goal at all, according to the authors. These findings show that in order to develop effective and
welfare-enhancing medicinal therapies, researchers must first examine how to effectively fulfil well-defined
social goals. When policies clearly consider the costs and benefits of developing population immunity, they
may help to strike the correct balance and increase social welfare.\footnote{The interested reader is invited to see \cite{1,2} and \cite{3} for a richer discussion.} 

In line with our ”Rule utilitarianism” approach mentioned earlier, a method borrowed from health economics
might be used to minimise the effect of pandemics on years of well-being by assessing the utility loss from
restrictions on economic activities in terms of QALYs. The value of gains in well-being over the course of
a year is equal for all persons, and the value of losses in well-being is likewise equal for all people when
this technique is used. This strategy, although common, is not without its drawbacks. On the basis of
the utilitarian principle and the premise that individuals have linear utility, it implicitly depends on this
approach. One dollar earned here equals one dollar gained somewhere else if utilities are linear (and same
across persons). This assumption makes it simple to replace monetary values everywhere. Even though
this classic utilitarian approach, using linear utility functions, is often employed in health care cost-effectiveness
analysis, it is clear that some kind of value judgement is made regarding the way in which expenses and benefits should be distributed.

 
\section{Conclusion}
 \label{sec:conclusion}

In this discussion, we focused on the context of the Italian reaction to the COVID-19 crisis in order to create a
useful resource for researchers and policymakers alike. Italy was the first country in Europe to be affected
by the COVID-19 outbreak and this is significant to our discussion for the following reason: whilst European
policymakers had the opportunity to learn from other countries’ responses and adapt their policies
accordingly, Italian policymakers – effectively being Europe’s ”patient zero” – did not have this relative
advantage. Therefore, \textit{rules} rather than \textit{data} asserted path dependence and informed decision-making in
the early stages of the crisis.

The most helpful ethical theories help to answer society’s questions, many of which have arisen due to the extraordinary challenges brought about by the pandemic. Classic utilitarianism, which employs the sum of utilities to evaluate
options, and average utilitarianism, which utilises average utility, are the two most well-known welfarist
principles. These concepts are often rejected since classical utilitarianism leads to some sub-optimal results,
while average utilitarianism occasionally deems the ceteris paribus addition of a person whose life is not
worth living to be beneficial. However, there are many principles that eschew both of these characteristics.
The central question of this paper was as follows: how should we think of citizens’ preferences? In a
democracy, difficult choices involving trade-offs of lives saved vs other economic and societal issues must
represent voters’ preferences. Is there a role for ethicists and philosophers in this process? Undoubtedly,
the answer is yes. A philosopher could assist policy choices by defining the moral principles that people are
ready to follow when difficult decisions must be made, and to advise politicians about the trade-offs that
would represent these preferences.

 Utilitarian welfare functions are appealing because they are consistent with the assumption that social preferences satisfy the von Neumann Morgenstern axioms \cite{H} and the Strong Pareto assumption\footnote{i.e, that society prefers one allocation over another if all individuals weakly prefer it and at least one individual strictly prefers it}. But
there are, of course, other ways to think about social welfare. For example, prioritarianism prioritises
outcomes with lower degrees of well-being by calculating a strictly increasing and concave transformation
of well-being. This prioritises those with lower degrees of well-being. Credited to moral philosopher Derek
Parfit \cite{P}, prioritarianism is a relatively new notion. In public policy, it is standard to define social welfare
as maximising a person’s expected present value of utility over the frequency with which he or she interacts
with others and over all time periods. An individual utility function is the aggregation of unit-comparable
functions. It is necessary to standardise individual utility functions so that transitory utility prior to the
COVID-19 pandemic is the same for all persons.

While utilitarianism is not without its detractors, utility theory appears to be well-equipped to deal with a number of theoretical challenges, such as questions around redistribution. As demonstrated in this paper, the axiomatic approach requires minimal conditions for normative plausibility with only one axiom sufficient to avoid both Repugnant and Very Sadistic Conclusions. 

Both the public policy and economic epidemiology literatures appear to be silent on the relationship between
axiomatic properties of social welfare relations and evaluating output costs of the lockdown. The aim of this
paper was to address this research gap and provide tentative suggestions for future research.

There exists a lacuna between two literatures, namely between economic epidemiology and population ethics. The critical evaluation of hidden assumptions does not go hand-in-hand with expertise in formal modelling.
As a result, important assumptions that underpin policy models are either incomplete or omitted. A concealed
assumption is not always incorrect; it may be perfectly appropriate. However, until we are aware that it
exists, we cannot assess it and so cannot reach a complete conclusion regarding the soundness of the model
or its output.

We therefore envisage avenues for further research where applied ethics and public policy are seen as close
complements, working towards a plausible, balanced, and consistent welfarist axiology that can rigorously
support models within economics and public health.

\newpage


\end{document}